\input{aipcheck}

\documentclass[
    ,final            
  ]
  {aipproc}

\layoutstyle{6x9}

\begin{document}

\title{Universal Extra Dimension models with right-handed neutrinos}


\author{Shigeki Matsumoto}{
  address={Institute for International Advanced Interdisciplinary Research,
     Tohoku University, Sendai, Miyagi 980-8578, Japan}
}

\author{Joe Sato}{
  address={Department of Physics, Saitama University, 
     Shimo-okubo, Sakura-ku, Saitama, 338-8570, Japan}
}

\author{Masato Senami}{
  address={ICRR, University of Tokyo, Kashiwa, Chiba 277-8582, Japan }
}

\author{Masato Yamanaka}{
  address={Department of Physics, Saitama University, 
     Shimo-okubo, Sakura-ku, Saitama, 338-8570, Japan} 
}

\begin{abstract}
 Relic abundance of dark matter is investigated in the framework of universal 
extra dimension (UED) models with right-handed neutrinos. These models are free from the KK graviton problem 
in the minimal UED model. The first KK particle of the right-handed neutrino is a dark matter candidate in this 
framework. When ordinary neutrino masses are large enough such as the degenerate mass spectrum case, the dark 
matter relic abundance can increase significantly. The scale of the extra dimension consistent with cosmological 
observations can be 500 GeV in the minimal setup of UED models with right-handed neutrinos.
\end{abstract}

\maketitle


\section{INTRODUCTION AND UED MODELS} 

  There is dark matter in the universe, and many models beyond the Standard Model (SM) have been proposed 
to explain the dark matter. Among those, Universal Extra Dimension (UED) models are one of interesting 
candidates for new physics, and it is worth investigating these models. We have solved problems inherent in 
these models, and calculated the allowed parameter by estimating the dark matter relic abundance. This 
proceeding is based on our works \cite{Matsumoto:2007dp, Matsumoto:2006bf}.

First, we briefly review UED model. This model is described in the five-dimensional 
space-time, where the extra dimension is compactified on an $S^1/Z_2$ orbifold with the radius $R$. As
a result of the compactification, many excited states of SM fields, called KK particles, appear. In this model,
the lightest KK particle (LKP) is stabilized by discrete symmetry, called KK parity. Thus, if the LKP is neutral, 
the LKP can be dark matter candidate.

Though the minimal UED model is good model, the model has two shortcomings. The first one is the KK 
graviton problem. In the parameter region where $1/R <$ 800 GeV, the KK graviton 
$G^{(1)}$ is the LKP, and the next LKP (NLKP) is the KK photon $\gamma^{(1)}$. Hence, $\gamma^{(1)}$ 
produced in the early universe decay into photons at late time universe, and these photons distort the CMB 
spectrum or the diffuse photon spectrum.

The second problem is the absence of neutrino masses. Since UED model has been constructed as minimal 
extension of the SM, neutrinos are treated as massless particles. However, neutrinos have mass. Therefore 
we must introduce neutrino masses into UED models.

\section{SOLVING THE PROBLEMS} 

In order to solve these problems, we introduce the right-handed neutrinos into UED models, and assume
that they form Dirac mass with ordinary neutrinos. The neutrino masses are expressed as 
$ \mathcal{L}_{\nu} = y_\nu \bar N L \Phi + {\rm h.c.} $, where $N$ is the right-handed neutrino, 
$L$ is the left-handed lepton. Then the second problem, the absence of the neutrino masses, are clearly 
solved. Once we introduce right-handed neutrinos into UED models, their KK particles automatically 
appear in the spectrum. The mass of the first KK right-handed neutrino, $N^{(1)}$, is estimated as
$m_{N^{(1)}}  \simeq \frac{1}{R} + O \left( \frac{m_{\nu }^2}{1/R} \right)~.$
After introducing the right-handed neutrinos, $N^{(1)}$ is the NLKP, and $\gamma^{(1)}$ is the next to next 
lightest KK particle.

The existence of the $N^{(1)}$ NLKP changes the late time decay of $\gamma^{(1)}$. In the models with the 
right-handed neutrino, $\gamma^{(1)}$ dominantly decays into $N^{(1)}$ and SM left-handed neutrino at 
tree level ($\gamma^{(1)}\rightarrow N^{(1)}\bar{\nu}$). In the KK neutral gauge boson mass matrix, mixing
angle is very small, and hence $\gamma^{(1)}$ can be regarded as KK U(1) gauge boson. Therefore $\gamma^{(1)}$ 
can decay into neutrinos through the hypercharge. On the other hand, the dominant 
decay mode associated with a photon is $\gamma^{(1)}\rightarrow G^{(1)}\gamma$. We estimeted the branching 
ratio of these decay modes
\begin{eqnarray}
 {\rm Br}
 =
 \frac{\Gamma(\gamma^{(1)} \rightarrow G^{(1)}\gamma) }{\Gamma(\gamma^{(1)} \rightarrow N^{(1)} \bar \nu)}
 =
 5 \times 10^{-7}
 \left(\frac{1/R}{500 {\rm GeV}}\right)^3
 \left(\frac{0.1 {\rm eV}}{m_\nu}\right)^2
 \left(\frac{\delta m}{1 {\rm GeV}}\right) ~.
\label{eq:br}
\end{eqnarray}
As a result, by introducing the right-handed neutrino into UED models, neutrino masses are introduced, and
problematic high energy photon emission is highly suppressed. Therefore, two problems in UED models have 
been solved simultaneously.

\section{$N^{(1)}$ DARK MATTER} 

$N^{(1)}$ cannot decay because it is forbidden by the kinematics. Since $N^{(1)}$ is neutral, massive, and 
stable, $N^{(1)}$ is also a dark matter candidate by introducing the right-handed neutrinos. 
Though abundance produced from decoupled $\gamma^{(1)}$ decay is same, abundance of $N^{(1)}$ dark 
matter produced from thermal bath is larger than that of $G^{(1)}$. Therefore, when $N^{(1)}$ is dominant 
component of dark matter, there are additional contribution to the dark matter relic abundance. 
As total dark matter number density becomes large, the dark matter mass allowed by cosmological 
observations, $\sim 1/R$, becomes small. Therefore, in order to determine the compactification scale $1/R$, 
we must evaluate the number density of $N^{(1)}$ dark matter.

Because the number density of the decoupled $\gamma^{(1)}$ has been calculated in previous works 
\cite{Kakizaki:2006dz}, we know 
$N^{(1)}$ number density produced from the decay of that. So, we calculate the $N^{(1)}$ 
number density produced from the thermal bath. To do this, we must include the thermal effects. In 
the thermal bath, particle mass receives thermal corrections. For example, the KK Higgs boson 
$\Phi^{(n)}$ mass is, 
\begin{eqnarray}
 m_{\Phi^{(n)}}^2(T)
 =
 m_{\Phi^{(n)}}^2(T=0)
 +
 \left[
  a (T) \cdot 3\lambda_h
  +
  x (T) \cdot 3 y_t^2
 \right] \frac{T^2}{12}~.
 \label{eq:thermalmass}
\end{eqnarray}
Coefficients $a(T)$ and $x(T)$ are counting factor how many KK modes contribute to the correction at the 
temperature $T$. Taking account of thermal masses, we calculated the $N^{(1)}$ number density.

\begin{figure}[h]
  \resizebox{30pc}{!}{\includegraphics{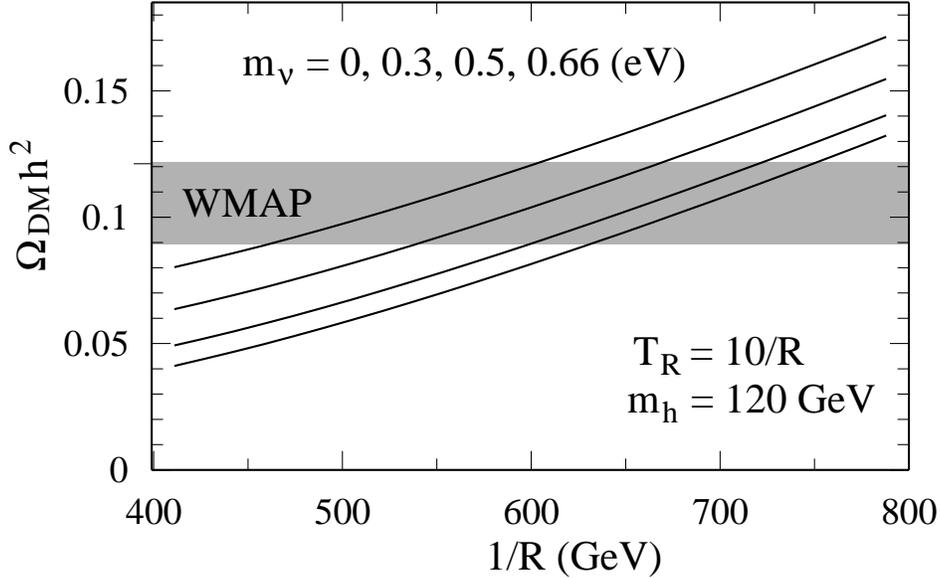}}
\caption{The dependence of the abundance on $m_\nu$ with fixed reheating 
          temperature $ T_R = 10 /R $ and $ m_h = 120 $ GeV. The solid lines
          are the relic abundance for $ m_\nu = 0, 0.3, 0.5, 0.66 $ eV from 
          bottom to top.  The gray band represents the allowed region from 
          the WMAP observation at the $2 \sigma$ level}
\end{figure}

Finally, we show the numerical result. Figure 1 shows the neutrino mass dependence of the abundance. The horizontal 
axis is the compactification scale $1/R$, and the vertical axis is the dark matter relic abundance. The 
lines correspond to the result with $ m_\nu = 0, 0.3, 0.5, 0.66 $eV from bottom to top. As shown in this 
figure, for large $m_\nu$, compactification scale can be less than 500 GeV. If compactification 
scale is less than 500 GeV, in ILC experiment, n=2 KK particles can be produced. n=2 KK particles are very 
important for discriminating UED from SUSY at collider experiment.

\section{SUMMARY} 

We have solved two problems in UED models (absence of the neutrino mass, KK graviton problem) by 
introducing the right-handed neutrinos. We have shown that by introducing right-handed neutrino, the dark 
matter is the KK right-handed neutrino $N^{(1)}$, and we have calculated the relic abundance of the $N^{(1)}$ 
dark matter. In our model, the compactification scale $1/R$ can be less than 
500 GeV. This fact has important consequence on the collider physics.



\bibliographystyle{aipproc}   

\bibliography{sample}

\IfFileExists{\jobname.bbl}{}
 {\typeout{}
  \typeout{******************************************}
  \typeout{** Please run "bibtex \jobname" to optain}
  \typeout{** the bibliography and then re-run LaTeX}
  \typeout{** twice to fix the references!}
  \typeout{******************************************}
  \typeout{}
 }


\end{document}